\documentclass[twocolumn]{svjour3}          
\smartqed  
\usepackage{graphicx}
\usepackage{amsfonts}
\usepackage{amssymb}
\usepackage{aas_macros}
\usepackage{color}
\newtheorem{exmp}{Example}

\begin{document}

\title{Black hole quasinormal modes in the era of LIGO
}


\author{Cecilia Chirenti}


\institute{C. Chirenti \at
              Centro de Matem\'atica, Computa\c c\~ao e Cogni\c c\~ao, UFABC, 09210-170 Santo Andr\'e-SP, Brazil\\
              Tel.: +55-11-49968299\\
              \email{cecilia.chirenti@ufabc.edu.br}           
           }

\date{Received: date / Accepted: date}

\maketitle

\begin{abstract}
After a long wait, gravitational wave astronomy has finally begun. Binary black hole mergers are being detected by LIGO and Virgo, and theorists are starting to receive a wealth of data to be analized. At this point we can at long last begin to test alternative theories of gravity and different models of compact objects. One powerful tool to do this is the perturbative analysis of background spacetimes. 
The objective of this brief review is to introduce the notion and analysis of black hole quasinormal modes, starting from the linear perturbation theory and including a brief discussion of numerical methods and astrophysical implications in the gravitational wave signals recently detected.
With these basic ingredients, more sophisticated analyses and applications are possible.
\end{abstract}

\section{Introduction}
\label{sec:intro}

Black holes are very simple objects. You might disagree with this statement; after all, classical black holes have singularities and event horizons, and one can also worry about open problems like the information paradox and firewalls. So, of course black holes can be very complicated! But still, black holes are very simple. Why? Because they are vacuum solutions, and therefore depend \emph{only} on the the spacetime geometry given by the Einstein field equations. 

Stellar (and cosmological) spacetimes have a matter content, and therefore a lot more complications. The nature of the dark energy and dark matter needed for the standard cosmological model is still a big mystery, and the equation of state of compact stars is also largely unknown. Black holes don't have any of that: they are pure geometry.

Let us take a look at the Schwarzschild solution:
\begin{equation}
ds^2 = -f(r)dt^2 + f(r)^{-1}dr^2 + r^2d\theta^2 + r^2\sin^2\theta d\phi^2\,,
\label{eq:Schwarzschild}
\end{equation}
with $f(r) = (1-2M/r)$. This solution describes a single, eternal, isolated and spherically symmetric black hole in vacuum. Thanks to Birkhoff's theorem, this solution is unique! And it is a very simple solution. There is no spin, no charge, no cosmological constant, no extra dimensions, etc. 

We will try to focus here mostly on \emph{real} astrophysical black holes. They are very likely to have spin, which we will discuss later. And they are very unlikely to be perfectly isolated. Their eternal rest can be disrupted by accretion, close passages of other astrophysical bodies, strong radiation or, even, by the ultimate fate of a binary system: the merger.

But let us not be concerned with such dramatic events for the time being. It is also fair to ask what will happen to a black hole when it is only slightly perturbed. Here a damped harmonic oscillator (or a bell, if you wish) is a good analogy. A perturbed black hole will go through the following stages:

\begin{enumerate}
\item transient
\item quasinormal mode ringdown
\item exponential or power law tail
\end{enumerate}

The transient depends on the details of the initial perturbation, but the ringdown shows characteristic frequencies of oscillation that encode information about the source (the black hole). The tail appears at late times, when the energy in the perturbation is too small to keep the oscillations going.

In the next sections we will see the main features of these quasinormal modes (Section \ref{sec:theory}), the properties of the gravitational quasinormal modes of the Schwarzschild black hole (Section \ref{sec:gravpert}), how they can be calculated (Section \ref{sec:methods}), an interesting interpretation (Section \ref{sec:interpretation}) and in which way they can be used to learn more from the recent observations of gravitational waves from binary black hole mergers \cite{PhysRevLett.116.061102,PhysRevLett.116.241103,PhysRevLett.118.221101,PhysRevLett.119.141101} (Section \ref{sec:observations}). Final remarks are presented in Section \ref{sec:final}.

This brief review is not intended to cover everything in this extensive topic, neither to provide a complete list of references, but it is thought as an introduction to the subject for advanced undergrad and graduate students. Therefore, I apologize in advance for the inevitable omissions. 

Much more detail and plenty of references can be found, for instance, in some excellent review papers \cite{1999LRR.....2....2K,1999CQGra..16R.159N,2011RvMP...83..793K}. Moreover, whenever I give examples from my own work, this is not intended as shameless self-advertising! Rather, this is a way to present the reader with examples from actual papers and non-trivial questions that they can try to answer.

\section{Linear perturbation theory and quasinormal modes}
\label{sec:theory}

The quasinormal modes that show up in the ringdown phase are characteristic modes of oscillation (that is, they are free oscillations) of a given spacetime, which means that they do not depend on the initial perturbations. 

Why are they called \emph{quasinormal} modes? The basic feature that makes them differ from \emph{normal} modes is that in this case the system is open, and loses energy through the emission of gravitational waves. This definition can be made much more mathematically precise (the quasinormal modes are the poles of the complex Green function), but it will do for now.

What exactly is oscillating with these quasinormal mode frequencies? If there is no matter content in a black hole spacetime, it must be spacetime itself that is oscillating. If we take the Einstein field equations
\begin{equation}
R_{\mu\nu} - \frac{1}{2}g_{\mu\nu}R = 8\pi GT_{\mu\nu}\,,
\label{eq:field}
\end{equation}
and take its trace, we find that
\begin{equation}
R = -8\pi GT\,.
\label{eq:trace}
\end{equation}
Now substituting eq.(\ref{eq:trace}) back in eq.(\ref{eq:field}), we have
\begin{equation}
R_{\mu\nu} = 8\pi G\left(T_{\mu\nu} -\frac{1}{2}g_{\mu\nu}T\right) \Rightarrow R_{\mu\nu} = 0\,,
\end{equation}
for a vacuum spacetime.

We can obtain a wave equation if we add a small perturbation $\delta g_{\mu\nu}$ to a fixed black hole background spacetime metric $g^0_{\mu\nu}$:
\begin{equation}
g_{\mu\nu} = g^0_{\mu\nu} + \delta g_{\mu\nu}\,,
\end{equation}
leading to the linearized Einstein field equations
\begin{equation}
\delta R_{\mu\nu} = 0\,,
\end{equation}
from which we can obtain a wave equation for the metric perturbations. Once we have the wave equation, it needs to be solved with the appropriate boundary conditions. In order to produce the characteristic, free oscillations of the black hole spacetime, the wave solution must be
\begin{itemize}
\item purely outgoing at infinity;
\item purely ingoing at the event horizon.
\end{itemize}

The radial part of the metric perturbations described by these oscillating wave solutions can be written as
\begin{equation}
\psi \propto e^{-i\omega t} = e^{-i(\omega_R + i\omega_I)t} = e^{\omega_I t}\cos(\omega_R t + \phi)\,,
\label{eq:sol}
\end{equation}
in terms of the complex frequency $\omega$, where $\omega_R$ is related to the period $T$ of the oscillation by
\begin{displaymath}
T = \frac{2\pi}{\omega_R}\,,
\end{displaymath}
and $\omega_I$ gives the characteristic timescale $\tau$ as
\begin{displaymath}
\tau = \frac{1}{\omega_I}\,.
\end{displaymath}

The sign of $\omega_I$ allows us to analyze the (linear) stability of the spacetime. As can be seen from eq. (\ref{eq:sol}), we have
\begin{itemize}
\item $\omega_I  < 0:$ exponential damping (stable);
\item $\omega_I > 0:$ exponential growth (unstable).
\end{itemize}

However, this has to be taken with some care. Note that our wave equation was obtained from the linearized Einstein equations, assuming that the perturbation $\delta g_{\mu\nu}$ in the metric is small when compared with the background metric. When $\omega_I > 0$, the perturbation grows exponentially, and the linear approximation breaks down. Therefore, this is an \emph{indication} that something is going on and that there might be an instability, but second order effects could play an important role and even make the solution stable\footnote{The fully nonlinear stability analysis of a spacetime is a much more complicated affair. So far, physicists have only succeeded to show that the Minkowski spacetime is stable (clearly a not so exciting result from the astrophysical point of view).}.

The frequencies of the quasinormal modes form a countable set of discrete frequencies. However, they do not form a complete set (they do not form a basis for any arbitrary perturbation), contrary to what usually happens with the normal modes of oscillation of a system.

We have discussed until now the perturbations in the metric of a black hole spacetime, that could be caused by some external perturbation. However a similar analysis can be done for a different type of perturbation: we can talk about the scattering of a classical field (scalar, electromagnetic, or other) in the fixed black hole background. Despite being a very different physical situation, the resulting wave equations are very similar, and this is why we talk about scalar, electromagnetic or gravitational perturbations of a given spacetime.

Below we solve the case of a scalar field as an example (since it is the shortest derivation!).

\begin{exmp}
Scattering of a scalar field on a \\Schwarzschild background
\end{exmp}
The Klein-Gordon equation governs the evolution of a scalar field, and it can be written as
\begin{equation}
\frac{1}{\sqrt{-g}}\partial_{\mu}(g^{\mu\nu}\sqrt{-g}\partial_{\nu}\psi) = 0\,.
\label{eq:KG}
\end{equation}
For the Schwarzschild metric (\ref{eq:Schwarzschild}), eq. (\ref{eq:KG}) becomes
\begin{eqnarray}
&-&f^{-1}\partial^2_t\psi + \frac{1}{r^2}[\partial_r(fr^2\partial_r \psi)] + \nonumber\\
&+&\frac{1}{r^2\sin\theta}[\partial_{\theta}(\sin\theta\partial_{\theta} \psi)] + 
\frac{1}{r^2\sin^2\theta}\partial^2_{\phi} \psi = 0\,.
\end{eqnarray}
Now we will proceed with the separation of variables as
\begin{equation}
\psi(t,r,\theta,\phi) = \sum_{\ell,m}e^{-i\omega t}Y_{\ell m}(\theta,\phi)\frac{\varphi(r)}{r}\,,
\end{equation}
which gives the following radial equation for $\varphi(r)$
\begin{equation}
\omega^2\varphi + \frac{f}{r}\left[\partial_r\left(fr^2\partial_r\left(\frac{\varphi}{r}\right)\right)\right]
-f\frac{\ell(\ell+1)}{r^2}\varphi = 0\,.
\label{eq:wavetemp}
\end{equation}
Now in order to get rid of the first order $r$-derivative, we will introduce the \emph{tortoise coordinate} $r_*$ by
\begin{equation}
\frac{dr_*}{dr} = \frac{1}{f}\,,
\end{equation}
which can be integrated to give%
\begin{equation}
r_* = r + 2M\ln(r-2M)\,.
\end{equation}
This new coordinate has the following limits:
\begin{equation}
r_* \to \infty\ \textrm{as}\ r \to \infty \quad \textrm{and} \quad r_* \to -\infty\ \textrm{as}\ r \to 2M\,.
\end{equation}
So the tortoise coordinate can be understood as a coordinate that stretches the spacetime outside of the black hole event horizon, as if it would approach the horizon \emph{very slowly}\ldots

Changing variables from $r$ to $r_*$ in eq.(\ref{eq:wavetemp}), the first derivative in the second term vanishes and we find
\begin{equation}
(\omega^2 + \partial^2_{r^*})\varphi = \left(1-\frac{2M}{r}\right)\left[\frac{\ell(\ell+1)}{r^2} + \frac{2M}{r^3}\right]\varphi\,.
_{\blacksquare} 
\label{eq:wavescalar}
\end{equation}

It can be shown that the ``effective potential'' multiplying $\varphi$ in the r.h.s. of eq.(\ref{eq:wavescalar}) can be written for different types of perturbations as \cite{1999LRR.....2....2K}
\begin{equation}
V_{\ell}(r) = \left(1-\frac{2M}{r}\right)\left[\frac{\ell(\ell+1)}{r^2} + \frac{2M(1-s^2)}{r^3}\right]\,,
\end{equation} 
where $s$ is the spin of the perturbation being considered:
\begin{eqnarray*}
\left\{
\begin{array}{l}
s = 0\ \textrm{for scalar perturbations,}\\
s = 1\ \textrm{for electromagnetic perturbations,}\\
s = 2\ \textrm{for axial gravitational perturbations,}
\end{array}
\right.
\end{eqnarray*}
and it is also worth noticing here that by symmetry we must have $\ell \ge 0$ for scalar perturbations, $\ell \ge 1$ for electromagnetic perturbations and $\ell \ge 2$ for gravitational perturbations.

\section{Gravitational perturbations of the Schwarzschild black hole}
\label{sec:gravpert}

The gravitational perturbations of the Schwarzschild black hole are described by two different equations of the form (\ref{eq:wavescalar}): the Regge-Wheeler equation \cite{1957PhRv..108.1063R} (for perturbations with axial parity) with
\begin{equation}
V^{\textrm{axial}}_{\ell}(r) = \left(1-\frac{2M}{r}\right)\left[\frac{\ell(\ell+1)}{r^2} - \frac{6M}{r^3}\right]\,,
\label{eq:RG}
\end{equation}
and the Zerilli equation (for perturbations with polar parity) \cite{1970PhRvD...2.2141Z} with 
\begin{eqnarray}
& &V^{\textrm{polar}}_{\ell}(r) = \frac{2}{r^3}\left(1-\frac{2m}{r}\right)\times \nonumber\\
&\times&\frac{9M^3 + 3c^2Mr^2 + c^2(1+c)r^3 + 9M^2cr}{(3M+cr)^2}\,,
\label{eq:Ze}
\end{eqnarray}
with $c = \ell(\ell+1)/2 - 1$.

These two potentials are clearly different, but their general features are similar. Let us analyze first the simpler Regge-Wheeler potential (\ref{eq:RG}). It represents a smooth potential barrier that goes to zero when $r \to 2M$ and when $r \to \infty$ \footnote{One could be tempted to say that $V^{\textrm{axial}}_{\ell}$ also has a zero at $r = 6M/[\ell(\ell+1)]$. However, this is not true: the largest possible value of $r$ (which will be given by $\ell = 2$) is $r = m$, that is, {\emph inside} the event horizon of the black hole!}. The Zerilli potential (\ref{eq:Ze}) has a similar structure.

It comes as a surprise that the Regge-Wheeler and the Zerilli equations have a very special property: they are isospectral \cite{1992mtbh.book.....C}. This means that they have the same spectrum of quasinormal mode frequencies. (This result is also valid for RNdS, but it is not valid for spacetimes with more than 4 dimensions \cite{2011RvMP...83..793K}.) We will sketch the proof for this unexpected result in the Example (\ref{ex:proof}) below.

However, before we do that, it is of special importance to discuss the physical interpretation of the two different parities (axial and polar) of the gravitational perturbations, that give rise to the two perturbation equations. We will start with the perturbed metric:
\begin{eqnarray}
ds^2 &=& -e^{2\nu}dt^2 + e^{2\mu_2}dr^2 + e^{2\mu_3}d\theta^2 + \nonumber\\
&+& e^{2\psi}(d\phi - \omega\, dt - q_2\,dr - q_3d\theta)^2\,,
\label{eq:pertmetric}
\end{eqnarray}
which describes a generic non-stationary asymmetric spacetime. We can restrict our study to axisymmetric perturbations because of the spherical symmetry of the Schwarzschild spacetime: any non-axisymmetric perturbations (with an $e^{im\phi}$ dependence) can be obtained from the axisymmetric perturbations by suitable rotations. 

The generic metric (\ref{eq:pertmetric}) gives the non-perturbed \\Schwarzschild metric if we chose
\begin{equation}
e^{2\nu} = e^{-\mu_2} = f\,, \quad e^{2\mu_3} = \frac{e^{2\psi}}{\sin^2\theta} = r^2
\end{equation}
and $\omega = q_2 = q_3 = 0$.

So it is clear that there are two different types of (gravitational) perturbations of the Schwarzschild metric:
\begin{enumerate}
\item the perturbation that gives small values to the metric coefficients that were zero ($\omega, q_2, q_3$): this perturbation induces frame dragging and imparts a rotation to the black hole; this is called the ``axial" perturbation.

\item the perturbation that gives small increments to the already non-zero metric coefficients \\($e^{2\nu},e^{2\mu_2},e^{2\mu_3},e^{2\psi}$): this is called the ``polar" perturbation.
\end{enumerate}
These two types of perturbations decouple (as expected) in the linearized Einstein field equations and lead to the Regge-Wheeler and Zerilli equations. Furthermore, axial perturbations transform as $(-1)^{\ell+1}$ under the parity operator, while polar perturbations transform as $(-1)^{\ell}$. This is why axial perturbations are also called odd perturbations, and polar perturbations are also called even perturbations.

\begin{exmp}
\label{ex:proof}
Proof sketch for isospectrality
\end{exmp}
Consider two wave equations of the form (\ref{eq:wavescalar}) with two potentials
\begin{equation}
 V^{\pm} = W^2(r_*) \pm \frac{dW(r_*)}{dr_*} + \beta\,,
\end{equation}
where $W(r_*)$ is a finite function and $\beta$ is a constant. If $\varphi^{\pm}(r_*)$ are solutions of the wave equations with $V^{\pm}(r_*)$, then it can be shown that
\begin{equation}
 \varphi^- \propto \left(W(r_*) - \frac{d}{dr_*}\right)\varphi^+(r_*)\,,
\end{equation}
Therefore $\varphi^+$ and $\varphi^-$ have the same eigenvalue $\omega$.$_{\blacksquare}$

\section{Methods for calculating quasinormal modes}
\label{sec:methods}

There are plenty of methods for solving a wave equation of the form (\ref{eq:wavescalar}), of course. Some methods that have been historically used for obtaining black hole quasinormal modes include:

\begin{itemize}
 \item shooting
 \item WKB
 \item characteristic integration
 \item continued fractions
 \item Frobenius series
 \item confluent Heun's equation
 \item Mashhoon method
\end{itemize}

I will not describe each one of those in detail here, they vary in complexity and accuracy and thorough expositions of all of them can be easily found in the literature.

We will see below as an example the semi-analytic Mashhoon method \cite{1984PhRvD..30..295F}, although it does not give such accurate results as some of the numerical methods listed above, because it is probably the simplest method to implement. It uses the P\"oschl-Teller potential (similarly to what is done with the Woods-Saxon potential in nuclear physics), to obtain an analytically treatable approximation of the black hole scattering potential.
\begin{exmp}
\label{ex:PT}
The Mashhoon method
\end{exmp}
The P\"oschl-Teller potential
\begin{equation}
 V_{PT} = \frac{V_0}{\cosh^2\alpha(r_* - r_*^0)}\,,
\end{equation}
has the same behavior we discussed above for the Regge-Wheeler potential, and it can be used as a good approximation to the effective potential $V(r_*,\alpha)$ in the black hole perturbation equations, after fitting $V_0$ and $r_*^0$. The advantage is that an analytical solution can be found, and it gives
\begin{equation}
 \omega = \pm \sqrt{V_0 - \frac{1}{4}\alpha^2} - i\alpha\left(n + \frac{1}{2}\right)\,, \quad n = 0,1,2,...
\end{equation}
From this solution we can see that the perturbations are stable (as we discussed above, since $\omega_I < 0$) and that the damping increases with the overtone number $n$. One drawback of this solution is that it gives a constant result for $\omega_R$, which is imprecise, as in reality $\omega_R$ also depends on $n$.$_{\blacksquare}$

\section{Physical interpretation of the asymptotically high overtones}
\label{sec:interpretation}

The asymptotically high overtones (with overtone number $n \to \infty$, or at least very large) of the gravitational Schwarzschild quasinormal modes have been found exactly by Motl \cite{2002gr.qc....12096M} and are given by
\begin{equation}
 \omega = \frac{\ln 3}{8\pi M} - \frac{i}{4M}\left(n + \frac{1}{2}\right)\,,
 \label{eq:high_omega}
\end{equation}
where we can see that now the real part is exactly constant (this was just an approximate result in the case of the Mashhoon method for small $n$) and the damping becomes arbitrarily large as $n$ increases. This expression can be used to explore a very interesting conjecture that links the quasinormal modes to black hole thermodynamics. If the mass of a black hole is quantized, it was suggested by Bekenstein and Mukhanov \cite{1995PhLB..360....7B} that this fact would lead to the quantization of the black hole area:
\begin{eqnarray}
 \Delta M &=& \hbar \Delta \omega \Rightarrow \Delta A = \Delta(4\pi R^2) = \Delta (16\pi M^2) = \nonumber\\
 &=& 32\pi M\Delta M = 32\pi \hbar M \Delta \omega\,.
\end{eqnarray}
This is very good as a conjecture but, what is the correct $\Delta \omega$ to be used? There is still no final answer to this question, and no confirmation that this is actually correct (that is why it is still just a conjecture). A first suggestion was made by Hod \cite{1998PhRvL..81.4293H}, by taking the real part of the asymptotic frequency (\ref{eq:high_omega}):
\begin{equation}
 \Delta \omega = \frac{\ln 3}{8\pi M}\,,
\end{equation}
but this choice presents some technical problems. Furthermore, this $\Delta \omega$ would represent a transition from the unperturbed state of the black hole to a state with very large $n$.

More recently, Maggiore suggested that the appropriate transition would be from a state with overtone number $n$ to a state with overtone number $n-1$ \cite{2008PhRvL.100n1301M}. This approach solves some of the problems posed by Hod's conjecture, and gives a different value for $\Delta \omega$, as we can show by considering a damped harmonic oscillator as a toy model:
\begin{equation}
 \ddot{\xi} + 2\gamma\dot{\xi} + \omega_0^2{\xi} = 0\,,
\end{equation}
which has the characteristic equation
\begin{equation}
 \omega^2 + 2i\gamma\omega - \omega_0^2 = 0\,,
\end{equation}
whose roots are the resonant frequencies of the system
\begin{equation}
 \omega = \pm\sqrt{\omega_0^2 - \gamma^2} - i\gamma\,,
\end{equation}
from which we can write 
\begin{equation}
 \omega_0^2 = \textrm{Re}(\omega)^2 + \textrm{Im}(\omega)^2\,.
\end{equation}
So for a $n \to n-1$ transition, one would get
\begin{equation}
 \Delta \omega_0 = \Delta \sqrt{\textrm{Re}(\omega)^2 + \textrm{Im}(\omega)^2} \approx \Delta\textrm{Im}(\omega) = \frac{1}{4M}\,,
\end{equation}
which finally gives
\begin{equation}
 \Delta A = 8\pi \hbar\,,
\end{equation}
or $\Delta A = 8\pi \ell_P$, after we recover the correct physical units.

All of this is still a conjecture but, if it can be proven to be true, it would provide a unique link between general relativity and quantum mechanics.
\footnote{Several challenges still have to be overcome. These conjectures lack universality when we consider black holes spactimes with electrical charge,  that are non-asymptotically flat or in extra dimensions \cite{Natario:2004jd}. An extension to the realm of greybody factors could provide a more direct route to a microscopic interpretation \cite{Harmark:2007jy}.}
 How could we explore such an amazing and promising idea? It would be great to make a detection that could provide evidence for this conjecture. However, only in the very distant future can we expect gravitational wave detectors capable of going beyond the fundamental quasinormal mode and maybe the first overtones. How would you try to prove (or disprove) the conjectured link between black hole quasinormal modes and black hole thermodynamics?

One counter-example is enough to kill a conjecture; proving it is a lot harder. Giving evidence that supports it is a way of making it more credible, but it is certainly not a proof. Some supporting evidence is given in example \ref{ex:nohorizon} below. 

\begin{exmp}
\label{ex:nohorizon}
Spacetimes without horizons
\end{exmp}
In \cite{Chirenti:2012ap} the following question was asked: if the asymptotic highly damped quasinormal modes are related to the quantum of the black hole horizon area, what should we expect in spacetimes with no horizons? 

These spacetimes can be regular stellar spacetimes or spacetimes containing naked singularities, like for instance the negative mass Schwarzschild singularity, the Reissner-Nordstr\"om naked singularity (a charged black hole with $Q > M$ or the Wyman's solution (a naked singularity arising from the minimal coupling of gravity to a charged massive scalar field).

It was shown that in such cases there are no asymptotic highly damped quasinormal modes. These solutions simply do not exist, when we take into account the other physical constraints. This result is not enough to prove the conjecture, but it certainly helps to support it! $_{\blacksquare}$

\section{Observation of quasinormal modes in the LIGO data}
\label{sec:observations}

The first gravitational quasinormal mode frequency of a Schwarzschild black hole corresponds to the fundamental ($n = 0$) quadrupole ($\ell = 2$) mode and it is
\begin{equation}
 M\omega = 0.37367 - 0.08896i\,,
\end{equation}
in geometrical units ($c = G = 1$) \cite{1999LRR.....2....2K}. For conversion to kHz, the frequencies must be multiplied by \\$(c^3/(GM_{\odot}))/(2\pi)\times(M_{\odot}/M)$. So, a hypothetical black hole of 1 solar mass would have a frequency $f = 12$ kHz and a damping time $\tau = 0.35$ ms. If a gravitational wave with these characteristics were to be detected, we would know for sure that it came from a 1 solar mass black hole. 

It is important to say that a similar perturbative analysis can also be performed for neutron stars. Neutron stars are expected to be the most compact astrophysical objects in nature (second only to black holes) with a typical mass of 1.4 $M_{\odot}$ and a typical radius of approximately 10-15 km. The boundary conditions for the perturbations have to be different in this case: still outgoing at infinity, but regular at the center of the star (since there is no horizon, the entire spacetime is accessible). 

Moreover, the quasinormal mode frequencies are going to depend on the equation of state of the star, which is still very much unknown in the extreme conditions that exist in the core of a neutron star\footnote{And this is not the whole story: neutron stars will have different \emph{families} of quasinormal modes, which will correspond to different restoring forces, like the pressure of the fluid, the buoyancy, the Coriolis force (for rotating stars), etc. Maybe now you will believe what I said in the beginning: black holes are simple!}. The frequency of the least damped mode of a neutron star (with a simple choice of polytropic equation of state) is given by \cite{Chirenti:2012wn}
\begin{eqnarray}
\label{eq:fstar}
f &=& 7.36 \times 10^{-2} + 55.80\sqrt{\frac{M}{R^3}}\,, \\
\frac{1}{\tau} &=& \frac{M^3}{R^4}\left[9.91 \times 10^{-2} - 0.33\left(\frac{M}{R} \right)\right]\,,
\label{eq:taustar}
\end{eqnarray}
so a typical neutron star with mass $M = 1.4M_{\odot}$ and radius $R = 14$ km will have a frequency $f = 1.6$ kHz
 and a damping time $\tau = 0.30$ s for the fundamental quasinormal mode. Therefore we can already see that black holes and neutron stars of comparable masses have very different signatures, and could be easily distinguished even if only the frequency of the first quasinormal mode is detected.

To infer the properties of the source (black hole, neutron star or other) from their gravitational wave signal is called the \emph{inverse problem}. We can see from eqs. (\ref{eq:fstar}) and (\ref{eq:taustar}) that, if $f$ and $\tau$ are observationally determined, the equations can be inverted to provide $M$ and $R$. These, in turn, would be very valuable to help in the determination of the equation of state of neutron stars.

In order to move now towards the gravitational wave observations recently announced by LIGO, we need to allow our black holes to have spin. The Kerr metric describes a spinning black hole, and perturbations of the Kerr black hole are given by the solutions of the Teukolsky equation \cite{1972PhRvL..29.1114T}. This perturbative analysis is similar to that we have described for the Schwarzschild black hole (but much more involved, of course!).

Using the P\"oschl-Teller potential, we can find an approximate solution for the gravitational quasinormal mode frequencies of the Kerr black hole:
\begin{equation}
 \omega \approx \frac{1}{3\sqrt{3}M}\left[\pm\left(\ell+\frac{1}{2}\right) + \frac{2am}{3\sqrt{3}M} - i\left(n+\frac{1}{2}\right)\right],
 \label{eq:Kerr}
\end{equation}
where $\ell \ge m\ \ge 1$ ($m$ is the azimuthal number) and $a \le M$ is the rotation parameter. Similarly to the inverse problem for neutron stars, we can already see here that, if both the frequency and the damping time of the first (fundamental) quasinormal mode can be detected, eq. (\ref{eq:Kerr}) can be used to obtain the mass $M$ and the rotation parameter $a$ of the black hole. 

The first LIGO observation of gravitational waves was announced on February 11th 2016, starting the age of gravitational wave astronomy. Since a long time before the first detection was made, it was believed that the first detections would come from the merger of compact binary systems. Generically, this type of gravitational wave signal can be divided in three parts: 1- inspiral, 2- merger and 3-ringdown. During most of the inspiral the binary components are far enough away from each other that they can be treated in the post-Newtonian approximation as point particles, whereas numerical relativity simulations must be performed to generate the signal expected during the merger. 

Based on the linear perturbation results, the ringdown portion of the signal may be used to discriminate between black holes and other possible sources. However, previous to the first detection, it was unclear whether the ringdown would be detectable with high enough amplitude to allow for this type of analysis. Actually, most people believed we wouldn't see the quasinormal modes in the ringdown at all in the first LIGO detections!

But the ringdown portion of the signal in the first gravitational wave detection event, GW150914, was strong enough that a direct fit of a damped sinusoid could be used to obtain the frequency and damping time of the fundamental (least damped) quasinormal mode \cite{PhysRevLett.116.221101}. The results indicated a Kerr black hole with $M = 62 M_{\odot}$ and $a = 0.68$. The rather heavy estimated masses of the binary components (36 $M_{\odot}$ and 29 $M_{\odot}$) and of the final black hole were the first surprise of the detections.

If we can detect in the future both the \emph{fundamental} quasinormal mode and the \emph{first overtone} in the ringdown, it will be possible to do more fundamental tests of general relativity: with both modes, it is possible to have two independent determinations of the final black hole's mass and spin. This can be used as a test of the Kerr metric and of the no-hair theorem.

With only the fundamental mode, it is already possible to test some exotic models, like some of the \emph{black hole alternatives} that have been proposed in the literature over the years. These alternative solutions have  different motivations, but in general they are very compact objects without an event horizon, that could only be distinguished from (ordinary) black holes by means of their gravitational radiation signature (at least as long as the surface of the object is at a small but not infinitesimal distance from the putative event horizon \cite{Cardoso:2016rao}). They include wormholes, boson stars, gravastars, superspinars, etc. 

In our last example \ref{ex:gravastar}, we will ask the question: what if GW150914 did not produce a black hole, but something else? Alternative models need to be ruled out one by one, and we see will what is the result for one of them.\footnote{Note, however, that the still large
uncertainties in our knowledge of the mass and angular momentum of the
resultant black hole does not currently allow us to completely discard some alternative
theories of gravity \cite{2016PhLB..756..350K}.} 

The \emph{gravastar} model \cite{Mazur:2004fk} provides an alternative scenario for the ultimate fate of the gravitational collapse of a massive star. As the star collapses and its radius shrinks,  it is conjectured that a phase transition would form
a de Sitter core inside the star.  As this “repulsive” core acts to stabilize the collapse, the  baryonic mass of the star forms  a  shell  of  stiff  matter  surrounding the core, creating an admittedly very exotic, although mathematically perfectly acceptable, compact object. This solution of the Einstein field equations could in principle be made with compactness arbitrarily close to that of a black hole. Due to its high surface redshift, the only possibly way to distinguish it from a black hole would be by their different spectrum of quasinormal modes \cite{Chirenti:2007mk}.

\begin{exmp}
\label{ex:gravastar}
The gravastar as a black hole alternative
\end{exmp}

How sure can we be that GW150914 created a black hole, and originated from a binary black hole system? This question was asked in \cite{Chirenti:2016hzd}, and a different scenario was proposed: suppose that the gravitational wave signal came from a binary gravastar system that inspiralled, merged and created a more massive gravastar. Would we be able to tell the difference in the signal? Or, rather, could we use the observed signal to rule out this possibility?

The gravastar solution discussed above can be described by the radius of the inner core, $r_1$, the radius of the external surface, $r_2$, and its total mass $M$. Alternatively to the two radii, we can use the compactness $\mu = M/r_2$ and the thickness of the shell $\delta = r_2 - r_1$ as parameters. Rotating gravastar solutions need to be constructed similarly to rotating stars, and the perturbative analysis also follows the same methods. 

After estimating the perturbative response of a rotating gravastar with $M = 62 M_{\odot}$ and $a = 0.68$ (with various values of $\mu$ and $\delta$) and comparing it with the LIGO data, it was possible to show that the fundamental quasinormal modes of these gravastars do not overlap with the damped sinusoidal fit of the data. The conclusion is, then, that GW150914 could not have created a gravastar (and therefore could not have originated as a gravastar binary system). $_{\blacksquare}$

\section{Final remarks}
\label{sec:final}
There are, by now, 4 confirmed (and one unconfirmed) gravitational wave detections from binary black holes mergers and one from a binary neutron star merger \cite{PhysRevLett.119.161101} that have already been announced. This is a truly exciting time to work with gravitational waves! We have already learned a great deal from the first detected signals, even though they have been nearly classic text book examples of what was predicted by the theory. For instance, we have found out that stellar mass black holes can be more massive than what was previously thought. 

Using linear perturbation theory and the new observations, we can finally put our mathematical analyses to work in a real astrophysical context. And it is beautiful to see the pieces coming together, and the actual physical reality (and detectability) of theoretical results that date back from the 1950's. 

We can only imagine how exciting it is going to be when we start detecting signals that are unexpected! This will be a marvellous opportunity to apply our physical knowledge, and today's students will be major players in this adventure. I hope to have given the reader, in this brief review, a short overview of the mathematical tools needed to treat the black hole quasinormal mode problem, the beginnings of the physical intuition to correctly interpret the results, and the motivation to get started.

\begin{acknowledgements}
This brief review is based on the lecture notes I prepared for a short course presented in the XVI Brazilian School of Cosmology and Gravitation (BSCG) held at the Brazilian Center for Physical Research (CBPF) in July 2017. 
I would like to thank Prof. Mario Novello for the invitation to participate in the XVI BSCG, the other lecturers and the students for a week of very interesting conversations and interactions, and CBPF for the financial support during my stay in Rio.
\end{acknowledgements}

\bibliographystyle{spphys}       
\bibliography{review}   

\end{document}